%% file: skeleton.tex
\documentclass[a4paper,11pt]{article}
\usepackage{pos}
\usepackage[justification=centering]{caption}

\title{Astronomy outreach in Namibia: \\ H.E.S.S. and beyond}
 \ShortTitle{Astronomy outreach in Namibia}

\author*[a,b]{Hannah Dalgleish}
\author[c]{Heike Prokoph}
\author[c]{Sylvia Zhu}
\author[b,d]{Michael Backes}
\author[a]{Garret Cotter}
\author[e]{Jacqueline Catalano}
\author[f]{Edna Ruiz-Velasco}
\author[b]{Eli Kasai}

\affiliation[a]{University of Oxford, Department of Physics, Oxford, UK}
\affiliation[b]{University of Namibia, Department of Physics, Windhoek, Namibia}
\affiliation[c]{Deutsches Elektronen-Synchrotron (DESY), Zeuthen, Germany}
\affiliation[d]{North-West University, Centre for Space Research, Potchefstroom, South Africa}
\affiliation[e]{ECAP, Friedrich-Alexander Universit\"{a}t Erlangen-N\"{u}rnberg, Germany}
\affiliation[f]{International Max Planck Research School for Astronomy and Cosmic Physics, University of Heidelberg}

\forColl{H.E.S.S.} 

\emailAdd{hannah.dalgleish@physics.ox.ac.uk}

\abstract{Astronomy plays a major role in the scientific landscape of Namibia. Because of its excellent sky conditions, Namibia is home to ground-based observatories like the High Energy Spectroscopic System (H.E.S.S.), in operation since 2002. Located near the Gamsberg mountain, H.E.S.S. performs groundbreaking science by detecting very-high-energy gamma rays from astronomical objects. The fascinating stories behind many of them are featured regularly in the ``Source of the Month'', a blog-like format intended for the general public with more than 170 features to date. In addition to other online communication via social media, H.E.S.S. outreach activities have been covered locally, e.g. through `open days' and guided tours on the H.E.S.S. site itself. An overview of the H.E.S.S. outreach activities are presented in this contribution, along with discussions relating to the current landscape of astronomy outreach and education in Namibia. There has also been significant activity in the country in recent months, whereby astronomy is being used to further sustainable development via human capacity-building. Finally, as we take into account the future prospects of radio astronomy in the country, momentum for a wider range of astrophysics research is clearly building --- this presents a great opportunity for the astronomy community to come together to capitalise on this movement and support astronomy outreach, with the overarching aim to advance sustainable development in Namibia.}

\FullConference{37$^{\rm{th}}$ International Cosmic Ray Conference (ICRC 2021)\\
		July 12th -- 23rd, 2021\\
		Online -- Berlin, Germany}

\begin{document}
\maketitle

\section{Astronomy outreach and capacity-building in Namibia}
The need for astronomy-related sciences is gaining increasing attention in Namibia, now recognised as a tool for fostering sustainable socio-economic growth. Most recently, in June 2021, the Ministry of Higher Education, Technology and Innovation launched the Space Science and Technology Policy\footnote{\url{https://africanews.space/namibia-lauches-national-space-science-and-technology-policy/}}. 
The desire to boost science, technology, and innovation via increased astronomy-related activity is similarly present on the rest of the continent, as illustrated by the African Union's African Space Strategy---For Social, Political and Economic Integration~\citep{au2019}.

\begin{figure}[ht]
    \centering
    \includegraphics[width=0.87\textwidth]{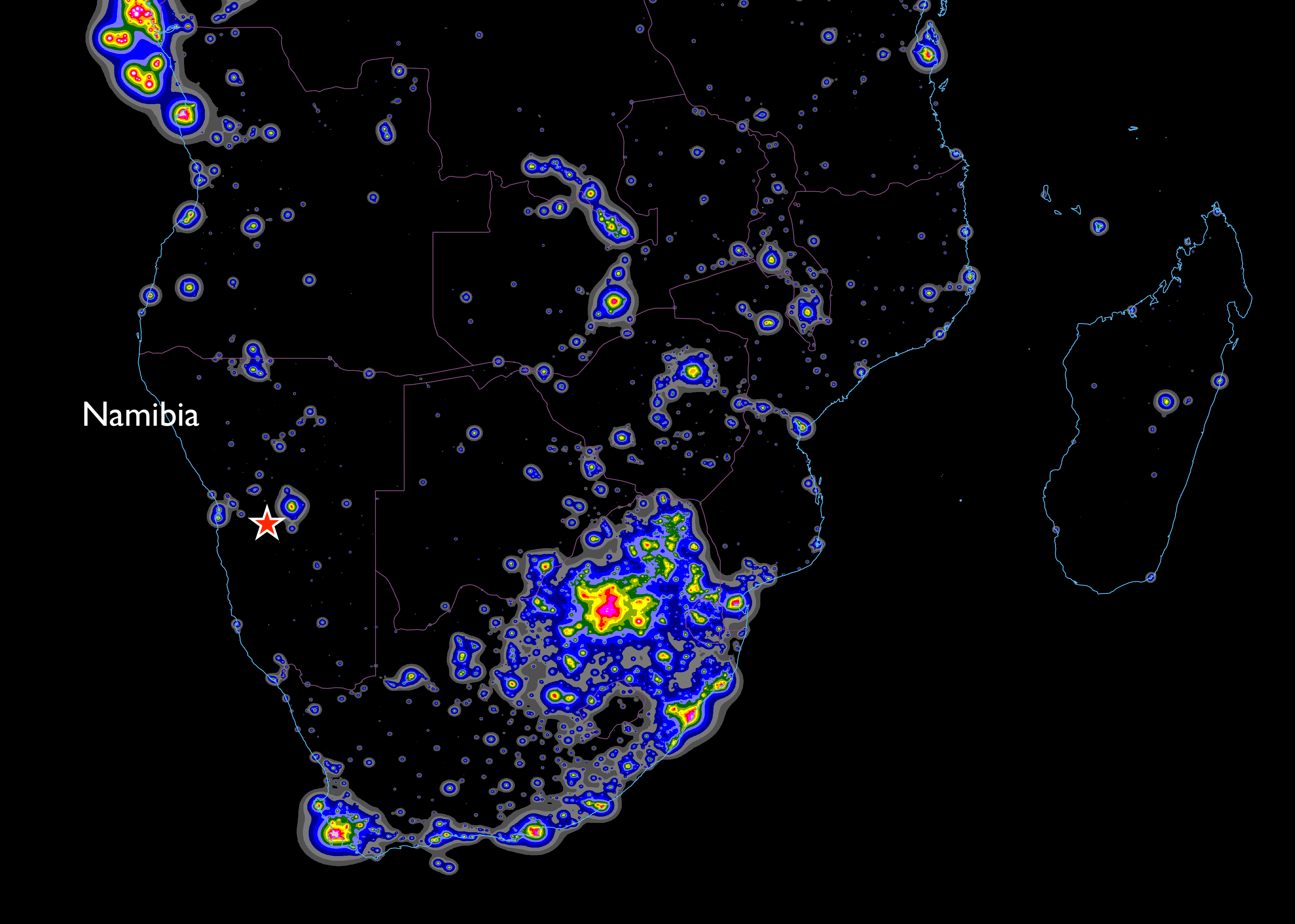}
    \includegraphics[width=0.105\textwidth]{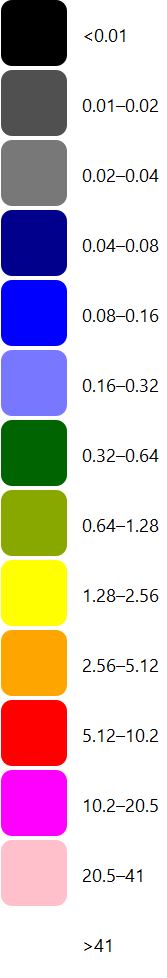}
    \caption{A map showing light pollution across Southern Africa, with Namibia in the West. The colour scale represents the ratio between artificial brightness and natural background sky brightness (assumed \\ to be 174 $\upmu$cd/m$^2$). The red star highlights the approximate location of the H.E.S.S. telescopes and the \\ Gamsberg mountain. Image courtesy of Fabio Falchi.}
    \label{fig:lp}
\end{figure}

Namibia also has a significant advantage in that the country is well suited for astronomical observations~\citep{backes2018}, given its very low levels of rainfall and light pollution (Fig.~\ref{fig:lp}; \citep{falchi2016}). As a result, Namibia is home to the High Energy Stereoscopic System (H.E.S.S.) telescopes~\citep{naurois2018}, as well as other current and planned observatories like the Africa Millimetre Telescope~(AMT)~\citep{backes2016:AMT}. Additionally, there have been significant efforts dedicated to building human capacity development in the long-term, through embracing Namibia's unique potential for astronomy-related activities. One of the main efforts has revolved around sustainable astrotourism, whereby indigenous star lore can be preserved, and tour guides can be trained in astronomical knowledge. Such training would thereby empower and equip local guides to give tours of the night sky, and ultimately increase and diversify their personal income as well as supporting rural communities as a whole~\citep{dalgleish2021a,dalgleish2021b}.

Another important program is being led by the AMT, which seeks to enhance astronomy outreach and education in Namibia through the use of a mobile planetarium and other educational tools. In 2019, the AMT joined forces with the Netherlands Research School for Astronomy (NOVA), Radboud University, the University of Namibia (UNAM), and the R\"{o}ssing Foundation, and spent a week travelling to schools in remote areas in northern and eastern Namibia. The mobile planetarium project now comes under the umbrella of the AMT's Social Impact Plan and will continue after Covid-19 restrictions are lifted, when the team intends to visit every school in the country over the next few years. 

\section{Astronomy outreach with H.E.S.S.}

Observatories provide a perfect setting for informal education~\citep{pompea2020}. As such, several research facilities around the world offer outreach programmes, including the Kitt Peak National Observatory in Arizona, the Southern African Large Telescope in South Africa, and the Thai National Observatory~\citep{isbell2003,govender2009,soonthornthum2015}.

Telescopes like H.E.S.S. also have a clear potential to be used for outreach purposes. Located in the Khomas highlands in Namibia, close to the Gamsberg mountain, the observatory is made up of five telescopes --- including the largest and most powerful Cherenkov telescope in the world. The H.E.S.S. telescopes work together to investigate cosmic gamma rays, and are used by an international collaboration of more than 230 scientists from 41 institutes in 15 countries. The instrument's name pays homage to Victor Franz Hess, who received the Nobel Prize in Physics in 1936 for his discovery of cosmic radiation.

\subsection{In-person outreach activities}
\label{outreach}
H.E.S.S. has offered a variety of outreach activities over the years. In 2012 an open day was held, and there were plans to host another one in 2020, which was inevitably postponed due to the pandemic. More recently, there have been tours made by VIPs, including UNAM's Vice Chancellor (July 2020), and the UK High Commissioner (May 2021). The H.E.S.S. site also welcomes visits from schools, tourists, and the general public --- the number of such visits is limited by a few key factors, described in further detail below.

Overall, the main purpose of H.E.S.S. is to deliver scientific research outputs, and thus, resources for outreach activities have been limited. Despite this, the local staff are always eager to host visitors and give tours during the day to those who are interested. 
With respect to Namibian schools, the majority are located too far away to be able to make a day trip to the telescopes, nor could they afford the accommodation and travel costs associated with visiting H.E.S.S. Travel remains an equal challenge for most of the schools in the local vicinity (i.e. Windhoek and Rehoboth) which do not have the funds available to hire a bus for the transportation of its pupils. Hence, it is primarily the private schools which have access to the resources needed to visit the observatory, but even then, it has been difficult to garner interest from this demographic, for reasons that are not currently clear.

In terms of the general public and international tourists, the main barrier is that H.E.S.S. is largely unknown to them. Although the observatory is listed in some German and French travel guides, it has been largely overlooked by the Namibia Tourism Board and other tourism organisations and operators. 

\subsection{Online communication and outreach}
In general, astronomers tend to focus their outreach efforts on traditional activities, like public lectures or radio interviews. However, virtual platforms can be equally effective for reaching members of the public to share new and exciting scientific developments. Although astronomers are highly communicative, only a small proportion are engaged in social media (less than 20\%;~\citep{entradas2019}), of which Facebook and Twitter are the most regularly used. Some astrophysicists are exceptionally popular on Twitter, such as Neil deGrasse Tyson (14.5M followers), Katie Mack (397K), and Sarafina Nance (125K) --- showcasing social media's capacity to inspire and engage the masses in astronomy research, and the scientists behind it.

The use of such online platforms have been taken up by large-scale astronomy-related organisations, including NASA (National Aeronautics and Space Administration), ESA (European Space Agency), ESO (European Southern Observatory), and the SKAO (Square Kilometre Array Observatory). Since as early as 2009, NASA has had a particularly unique engagement with its followers on social media --- through the NASA Social program (formerly known as Tweetups) --- whereby selected candidates are invited to go behind-the-scenes to meet with NASA personnel, and even to attend Space shuttle launches~\citep{vaneperen2011}. 
Another unique example of online communication comes from ESA, which successfully executed a high-impact online public engagement campaign for the  \textit{Rosetta} mission via blogging, Twitter, Facebook, YouTube, and more --- reaching millions of people worldwide~\citep{baldwin2016}. 

Some astronomy collaborations also maintain an online presence, including H.E.S.S, MAGIC (Major Atmospheric Gamma Imaging Cherenkov), SDSS (Sloan Digital Sky Survey), 
and the EHT (Event Horizon Telescope). Looking at H.E.S.S. more closely, outreach activities were initiated by the Source of the Month\footnote{\url{https://www.mpi-hd.mpg.de/hfm/HESS/pages/home/som/}}, a blog written by members of the collaboration which features an interesting gamma-ray object each month. The first blog was published in October 2004, and there have been more than 170 features since (there was a hiatus between October 2013--December 2015). 

Alongside the H.E.S.S. website, social media accounts are used to share notable science results, such as the recent discovery of the gamma-ray burst (GRB 190829A), the most energetic radiation and longest gamma-ray afterglow observed thus far~\citep{hess2021}. These posts are often interacted with the most, especially when they include a video or GIF. Other posts promote the Source of the Month, share the achievements and contributions of collaboration members, celebrate PhD defences, and present the H.E.S.S. prize winners. The YouTube account also features short ($\sim$4 minute) interviews with on-site staff. A summary of the H.E.S.S. social media accounts with relevant statistics is shown in Table~\ref{tab:socialmedia}.

\begin{table}[ht]
\centering
\begin{tabular}{lcccc}
Platform 	& Account created 	& Total followers & Total posts & Total views  \\ 
\hline \hline
Twitter 	& November 2010 	& 1222 	& 647 	& $\sim570$k$^\dag$\\ 
Facebook 	& March 2016  		& 1705 	& $\sim90$	& $\sim22.6$k \\ 
YouTube 	& September 2020 	& 34 	& 3 	& 586 \\ 
Instagram	& November 2020 	& 277 	& 18 	& 532$^*$ \\ 
\end{tabular}
\caption{Overview of the H.E.S.S. social media accounts. Data collected on 29th June 2021.\\
$^{\dag}$ This value refers to the total number of Twitter impressions since September 2014. \\
$^*$ This value refers to total likes across all posts instead of views.}
\label{tab:socialmedia}
\end{table}

\section{Future prospects}
At the most recent H.E.S.S. collaboration meeting in April 2021, talks and breakout discussions specifically focused on outreach and social impact occurred for the first time. The desire to boost astronomy outreach at H.E.S.S. was strongly apparent, where the discussion revolved around overcoming the obstacles described in Section~\ref{outreach}. Additional funds would open up several possibilities, such as supporting local schools to visit the observatory, or hiring someone locally who could coordinate visits via tour operating companies. Such a staff member could also support public engagement on the aforementioned online channels. To boost the social impact of H.E.S.S. even further, the collaboration could consider implementing a teacher training program, or even hosting an artist-in-residence. 

Ultimately, this discourse comes at a perfect time. H.E.S.S. has a fundamental role to play in the Namibian vision for stimulating the next generation of scientists, while building momentum for human capacity development through astronomy~\citep{dalgleish2020}. In Namibia and neighbouring countries, societal impact plans are already underway for two new observatories, namely the AMT and the Square Kilometre Array Observatory~\citep{ska2021}. Together, these telescopes will not only advance our understanding of the Universe, but can help to stimulate sustainable development in Southern Africa. 

\acknowledgments
We are especially grateful to Fabio Falchi for creating the light pollution map of Southern Africa (Figure~\ref{fig:lp}), and to Stefan Wagner for his very helpful comments.

The support of the Namibian authorities and of the University of Namibia in facilitating the construction and operation of H.E.S.S. is gratefully acknowledged, as is the support by the German Ministry for Education and Research (BMBF), the Max Planck Society, the German Research Foundation (DFG), the Helmholtz Association, the Alexander von Humboldt Foundation, the French Ministry of Higher Education, Research and Innovation, the Centre National de la Recherche Scientifique (CNRS/IN2P3 and CNRS/INSU), the Commissariat \`{a} l'\'{e}nergie atomique et aux \'{e}nergies alternatives (CEA), the U.K. Science and Technology Facilities Council (STFC), the Knut and Alice Wallenberg Foundation, the National Science Centre, Poland grant no. 2016/22/M/ST9/00382, the South African Department of Science and Technology and National Research Foundation, the University of Namibia, the National Commission on Research, Science \& Technology of Namibia (NCRST), the Austrian Federal Ministry of Education, Science and Research and the Austrian Science Fund (FWF), the Australian Research Council (ARC), the Japan Society for the Promotion of Science and by the University of Amsterdam.

We appreciate the excellent work of the technical support staff in Berlin, Zeuthen, Heidelberg, Palaiseau, Paris, Saclay, T\"{u}bingen and in Namibia in the construction and operation of the equipment. This work benefited from services provided by the H.E.S.S. Virtual Organisation, supported by the national resource providers of the EGI Federation.

We also acknowledge support from the UKRI STFC Global Challenges Research Fund project ST/S002952/1 and Exeter College, Oxford.

\clearpage
\section*{Full Authors List: \Coll\ Collaboration}
%
%
\scriptsize
\noindent
\input{authors_all_ICRC}
%

\end{document}

%% file: ICRC2021 template/authors_ALL_ICRC.tex
\scriptsize
\noindent
H.~Abdalla$^{1}$, 
F.~Aharonian$^{2,3,4}$, 
F.~Ait~Benkhali$^{3}$, 
E.O.~Ang\"uner$^{5}$, 
C.~Arcaro$^{6}$, 
C.~Armand$^{7}$, 
T.~Armstrong$^{8}$, 
H.~Ashkar$^{9}$, 
M.~Backes$^{1,6}$, 
V.~Baghmanyan$^{10}$
% ...

\medskip

\noindent
$^{1}$University of Namibia, Department of Physics, Private Bag 13301, Windhoek 10005, Namibia\\
$^{2}$Dublin Institute for Advanced Studies, 31 Fitzwilliam Place, Dublin 2, Ireland\\
$^{3}$Max-Planck-Institut f\"ur Kernphysik, P.O. Box 103980, D 69029 Heidelberg, Germany\\
$^{4}$High Energy Astrophysics Laboratory, RAU,  123 Hovsep Emin St  Yerevan 0051, Armenia\\
$^{5}$Aix Marseille Universit\'e, CNRS/IN2P3, CPPM, Marseille, France\\
$^{6}$Centre for Space Research, North-West University, Potchefstroom 2520, South Africa\\
$^{7}$Laboratoire d'Annecy de Physique des Particules, Univ. Grenoble Alpes, Univ. Savoie Mont Blanc, CNRS, LAPP, 74000 Annecy, France\\
$^{8}$University of Oxford, Department of Physics, Denys Wilkinson Building, Keble Road, Oxford OX1 3RH, UK\\
$^{9}$IRFU, CEA, Universit\'e Paris-Saclay, F-91191 Gif-sur-Yvette, France\\
$^{10}$Instytut Fizyki J\c{a}drowej PAN, ul. Radzikowskiego 152, 31-342 Krak{\'o}w, Poland\\
% ...
